\documentclass[twocolumn,showpacs,amsmath,amssymb,superscriptaddress,floatfix]{revtex4}
\usepackage{graphicx}
\usepackage{dcolumn}
\usepackage{bm}
\usepackage{verbatim}
\newcommand{\dd}{\mbox{\textrm{d}}}

\begin{document}
\title{Study of the $\boldsymbol{pd(dp) \to{} ^3\textrm{He}\,\pi\pi}$ reactions close to threshold}

\author{F.~Bellemann}%
\affiliation{Institut f\"ur Strahlen- und Kernphysik, Universit\"at Bonn, D-53115 Bonn, Germany}
\author{A.~Berg}
\affiliation{Institut f\"ur Strahlen- und Kernphysik, Universit\"at Bonn, D-53115 Bonn, Germany}
\author{J.~Bisplinghoff}%
\affiliation{Institut f\"ur Strahlen- und Kernphysik, Universit\"at Bonn, D-53115 Bonn, Germany}
\author{G.~Bohlscheid}%
\affiliation{Institut f\"ur Strahlen- und Kernphysik, Universit\"at Bonn, D-53115 Bonn, Germany}
\author{J.~Ernst}
\affiliation{Institut f\"ur Strahlen- und Kernphysik, Universit\"at Bonn, D-53115 Bonn, Germany}
\author{C.~Henrich}
\affiliation{Institut f\"ur Strahlen- und Kernphysik, Universit\"at Bonn, D-53115 Bonn, Germany}
\author{F.~Hinterberger}
\affiliation{Institut f\"ur Strahlen- und Kernphysik, Universit\"at Bonn, D-53115 Bonn, Germany}
\author{R.~Ibald}
\affiliation{Institut f\"ur Strahlen- und Kernphysik, Universit\"at Bonn, D-53115 Bonn, Germany}
\author{R.~Jahn} \email[E-mail: ]{jahn@hiskp.uni-bonn.de}%
\affiliation{Institut f\"ur Strahlen- und Kernphysik, Universit\"at Bonn, D-53115 Bonn, Germany}
%
%
\author{R.~Joosten}
\affiliation{Institut f\"ur Strahlen- und Kernphysik, Universit\"at Bonn, D-53115 Bonn, Germany}
\author{K.~Kilian}
\affiliation{Institut f\"ur Kernphysik, Forschungszentrum J\"ulich, D-52425 J\"ulich, Germany}
\author{A.~Kozela}
\affiliation{Institute of Nuclear Physics, PL-31342 Krak\'ow, Poland}
\author{H.~Machner}
\affiliation{Fakult\"at f\"ur Physik, Universit\"at Duisburg-Essen,
Lotharstrasse 1, D-47048 Duisburg, Germany}
\author{A.~Magiera}
\affiliation{Institute of Physics, Jagellonian University, PL-30059 Krak\'ow, Poland}
%
%
%
%
\author{J.~Munkel}
\affiliation{Institut f\"ur Strahlen- und Kernphysik, Universit\"at Bonn, D-53115 Bonn, Germany}
\author{P.~von~Neumann-Cosel}
\affiliation{Institut f\"ur Kernphysik, Technische Universit\"at Darmstadt, D-64289 Darmstadt, Germany}
%
%
\author{P.~von Rossen}
\affiliation{Institut f\"ur Kernphysik, Forschungszentrum J\"ulich, D-52425 J\"ulich, Germany}
\author{H.~Schnitker}
\affiliation{Institut f\"ur Strahlen- und Kernphysik, Universit\"at Bonn, D-53115 Bonn, Germany}
\author{K.~Scho}
\affiliation{Institut f\"ur Strahlen- und Kernphysik, Universit\"at Bonn, D-53115 Bonn, Germany}
\author{J.~Smyrski}
\affiliation{Institute of Physics, Jagellonian University, PL-30059 Krak\'ow, Poland}
%
%
\author{R.~T\"olle}
\affiliation{Institut f\"ur Kernphysik, Forschungszentrum J\"ulich, D-52425 J\"ulich, Germany}
\author{C.~Wilkin}
\affiliation{Physics and Astronomy Department, UCL, London WC1E 6BT, United Kingdom}
\collaboration{The COSY-MOMO collaboration} \noaffiliation

\date{\today}

\vspace*{1cm}
%
\begin{abstract}
New experimental data on the $pd\to{} ^3\textrm{He}\,\pi^+\pi^-$ reaction
obtained with the COSY-MOMO detector below the three-pion threshold are
presented. The reaction was also studied in inverse kinematics with a
deuteron beam and the higher counting rates achieved were especially
important at low excess energies. The comparison of these data with inclusive
$pd\to{} ^3\textrm{He}\,X^0$ rates allowed estimates also to be made of
$\pi^0\pi^0$ production. The results confirm our earlier findings that close
to threshold there is no enhancement at low excitation energies in the
$\pi^+\pi^-$ system, where the data seem largely suppressed compared to phase
space. Possible explanations for this behavior, such as strong $p$ waves in
the $\pi^+\pi^-$ system or the influence of two-step processes, are explored.
\end{abstract}

\maketitle

%
%
\section{Introduction}
\label{Introduction}

The ABC effect is an enhancement of the two-pion invariant mass
($M_{\pi\pi}$) spectrum close to threshold that has been observed in certain
nuclear reactions. It manifests itself through a peak at a mass of about
310~MeV/$c^2$ with a width $\approx 50$~MeV/$c^2$. However, these values
change with experimental conditions and there is much evidence to show that
the ABC is a kinematic effect, associated with the presence of nucleons,
rather than being a genuine $s$-wave $\pi\pi$ resonance~\cite{AMS2012}.

The effect was first identified by Abashian, Booth, and Crowe (ABC) in
measurements of the inclusive cross sections for $pd \to{}^3\textrm{He}\,X^0$
at a beam energy of $T_p=743$~MeV~\cite{ABA1963}. The lack of a similar
signal in the $pd \to{}^3\textrm{H}\,X^+$  case shows that the effect has to
be dominantly in the $\pi\pi$ isospin $I_{\pi\pi}=0$ channel. Apart from
phase space effects, one would then expect that the $\pi^+\pi^-$ component in
the production of the ABC should be twice as strong as the $\pi^0\pi^0$.

The original ABC data covered only production of the $^3$He in the forward
hemisphere with respect to the proton beam direction in the center-of-mass
system (CMS)~\cite{ABA1963}. By using a deuteron beam with an energy about
twice as high, the acceptance was increased significantly and allowed the ABC
effect to be observed inclusively in both hemispheres at
Saclay~\cite{BAN1973}.

In order to discuss data in different kinematic regions, it is convenient to
label them in terms of the excess energy $Q=W-M_{^3\rm{He}}-2M_{\pi}$, where
$W$ is the total CMS energy. The original inclusive ABC data were obtained at
$Q=184$~MeV with respect to the charged pion threshold~\cite{ABA1963}.
Exclusive measurements of both the $pd \to{} ^3\textrm{He}\,\pi^+\pi^-$ and
$pd  \to{} ^3\textrm{He}\,\pi^0\pi^0$ differential cross sections were
carried out at the even higher excess energy of $Q=269$~MeV by the
CELSIUS-WASA collaboration~\cite{BAS2006} and these were complemented by
later measurements of $pd  \to{} ^3\textrm{He}\,\pi^0\pi^0$ at $Q=338$~MeV by
the WASA collaboration at COSY~\cite{ADL2015,RIO2014}. The data supported the
conclusion that at low $\pi\pi$ invariant mass $M_{\pi\pi}$ the ABC effect
was of dominantly isoscalar ($I_{\pi\pi}=0$) nature, though corrections had
to be made to account for the pion mass differences. However, the charged
pion data suggested that there could be some $I_{\pi\pi}=1$ contribution at
large $M_{\pi\pi}$. When the kinematics of the full three-body final state
were reconstructed, the exclusive experiments also allowed the distributions
in the $\pi{}^3$He invariant mass to be evaluated. These seemed to show some
reflections of the $\Delta(1232)$ distribution.

Although the systematics were less well controlled, much higher statistics on
the $dp \to{} ^3\textrm{He}\,\pi^+\pi^-$ reaction at a similar excess energy
were obtained by the COSY-ANKE collaboration by using a deuteron beam
incident on a hydrogen target~\cite{MIE2016}. The difference between the
$\pi^+{}^3$He and $\pi^-{}^3$He invariant mass distributions was an
indication of some interference between $I_{\pi\pi}=1$ and $I_{\pi\pi}=0$
amplitudes. It should be noted that, although the set-ups of the CELSIUS-WASA
and ANKE experiments were very different, both sets of measurements were
carried out in the forward CM hemisphere between the incident proton and
final $^3$He.

Both the WASA and the ANKE experiment show the importance of the
$\Delta(1232)$ in two-pion production at high excess energies and so it is
not unexpected that the results could look rather different at low $Q$,
\emph{i.e.}, below the threshold for $\Delta$ production. Nevertheless, there
was surprise when the first exclusive $pd \to{} ^3\textrm{He}\,\pi^+\pi^-$
results emerged from the COSY-MOMO collaboration; these showed that at
$Q=70$~MeV there was no sign of any ABC effect~\cite{BEL1999}. The data were
low compared to phase space at small $M_{\pi\pi}$ and, indeed, they could be
modeled as if there were a $p$-wave between the $\pi^+\pi^-$ pair. No
comparison of the $\pi^{+}\,^3$He and $\pi^{-}\,^3$He invariant mass
distributions could be made because, in the absence of a magnetic field in
the MOMO detector, it was not possible to distinguish the sign of the charge
on an individual pion. However, the absence of an ABC effect even closer to
threshold was confirmed in low-statistics data obtained at $Q=28$~MeV at
CELSIUS~\cite{AND2000}. The different behavior between low and high $Q$ was
also noted in the quasi-free production reaction $dd  \to{} n_{\rm sp}{}
^3\textrm{He}\,\pi^0\pi^0$, where $n_{\rm sp}$ is a \emph{spectator} neutron
that was reconstructed from the measurements of the $^3$He and two neutral
pions in the WASA detector~\cite{ADL2015,RIO2014}. The Fermi motion in the
deuteron allowed the authors to estimate the cross section for $pd  \to{}
^3\textrm{He}\, \pi^0\pi^0$ over a range of values of $Q$.

In view of the marked differences between the observations for large and
small excess energy, it was decided to carry out further measurements with
the MOMO detector at excess energies above and below our previous value of
$Q=70$~MeV. The experimental set-up, with the $^3$He being measured in a high
resolution spectrograph and the charged pions the MOMO detector, is described
in some detail in Sec.~\ref{setup}. One conclusion that is evident from this
discussion is that the acceptance for the  $dp \to{}
^3\textrm{He}\,\pi^+\pi^-$ reaction with a deuteron beam is significantly
higher than that with incident protons. The doubling of the incident momenta
leads to generally faster particles that are pushed into smaller angular
regions. The gain by using a deuteron beam is especially important at low
excess energy $Q$ because the cross section falls very rapidly as threshold
is approached.

Data taken in $pd$ kinematics  are first presented at an excess energy of $Q
= 92$~MeV to investigate the anti-ABC effect first noted in the MOMO 70~MeV
results~\cite{BEL1999}. Estimates of the cross sections for $pd \to{}
^3\textrm{He}\,\pi^0\pi^0$ were also made in both these cases by comparing
the data sets obtained with and without the $\pi^+\pi^-$ detection in MOMO.
The comparison of charged and neutral pion data indicates that there must be
a very significant fraction of $I_{\pi\pi}=1$ production at these energies.
This is consistent with the direct measurements of the $pd \to{}
^3\textrm{He}\,\pi^+\pi^-$ and $pd \to{} ^3\textrm{He}\,\pi^0\pi^0$ cross
sections carried out at CELSIUS at $Q=28$~MeV~\cite{AND2000}. This energy was
repeated with higher statistics at MOMO in $pd$ kinematics~\cite{BOH1998}
before being investigated fully with a deuteron beam. The consistency of the
MOMO $pd$ and $dp$ data at $Q\approx 28$~MeV gives confidence in the
acceptance estimates in the analysis. This allowed data to be taken with a
deuteron beam at $Q = 8$~MeV, which would have been highly problematic in
$pd$ kinematics. The results of these measurements are reported in
Sec.~\ref{results}.

Though a suppression of the data at low $M_{\pi\pi}$ might be a signal for
$p$-wave pion pairs, there are other possibilities, as discussed in
Sec.~\ref{interpretation}. In a two-step model the reaction is closely linked
to that for $\pi^-p\to \pi^0\pi^0n$~\cite{FAL2000}, where the sum of a
contact term and production via the Roper resonance can also deplete the
cross section near the $\pi\pi$ threshold~\cite{VAC1999}. Our conclusions and
suggestions for further work are presented in Sec.~\ref{conclusions}.

%
%
\section{Experimental set-up}
\label{setup}

The layout of the experimental setup with the MOMO (Monitor Of Mesonic
Observables) detector was described in our previous publication that reported
the $pd \to{} ^3\textrm{He}\,K^+K^-$ measurements~\cite{BEL2007}. An external
proton or deuteron beam from the COSY accelerator of the Forschungszentrum
J\"{u}lich was incident on a 4~mm thick liquid deuterium or hydrogen target
with 1.5~$\mu$m mylar windows~\cite{JAE1994}. A beam diameter of less than
2~mm led to precise determination of the emission angles.

\begin{figure}[hbt]
\begin{center}
\includegraphics[width=0.48\columnwidth]{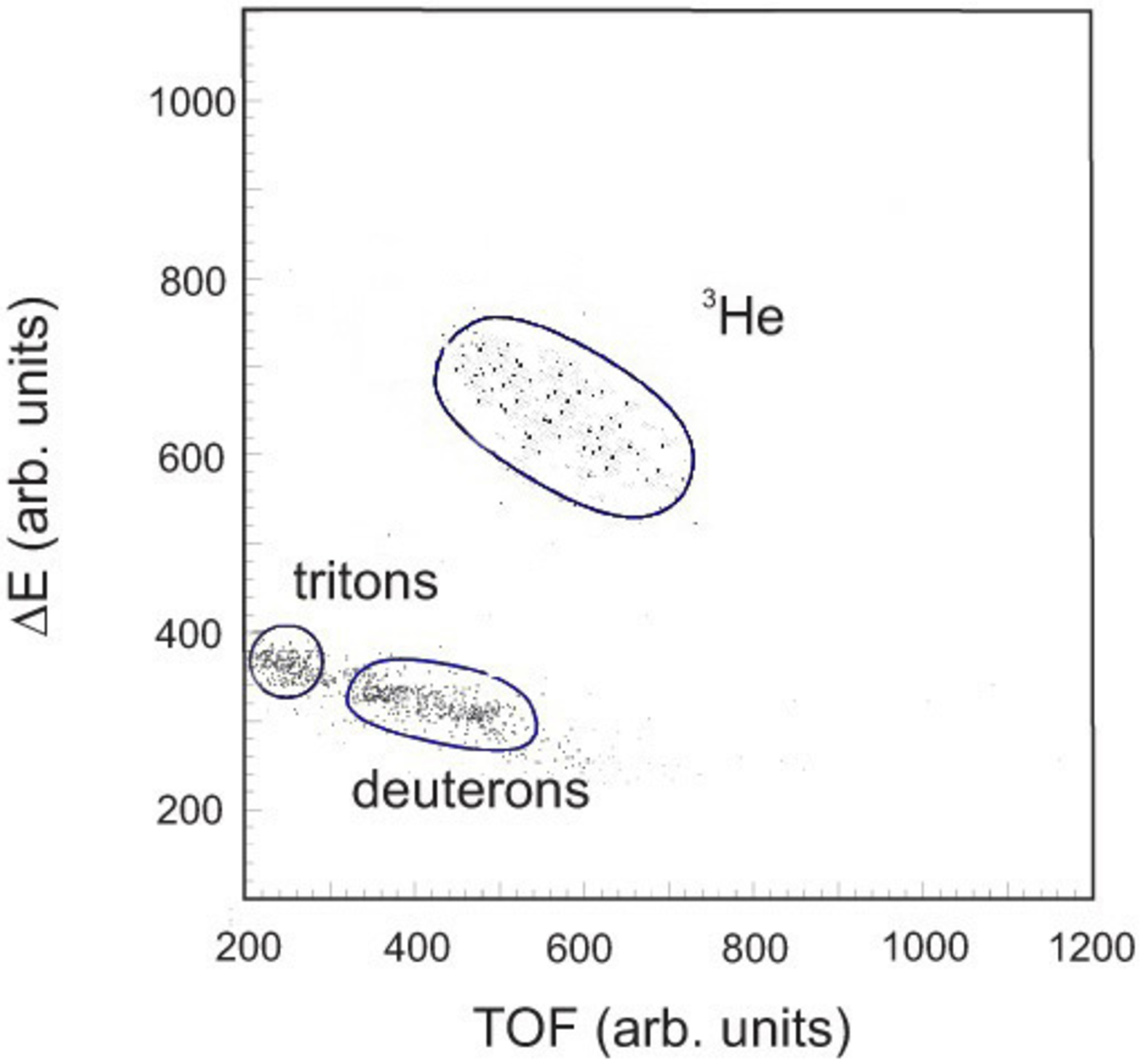}
\includegraphics[width=0.50\columnwidth]{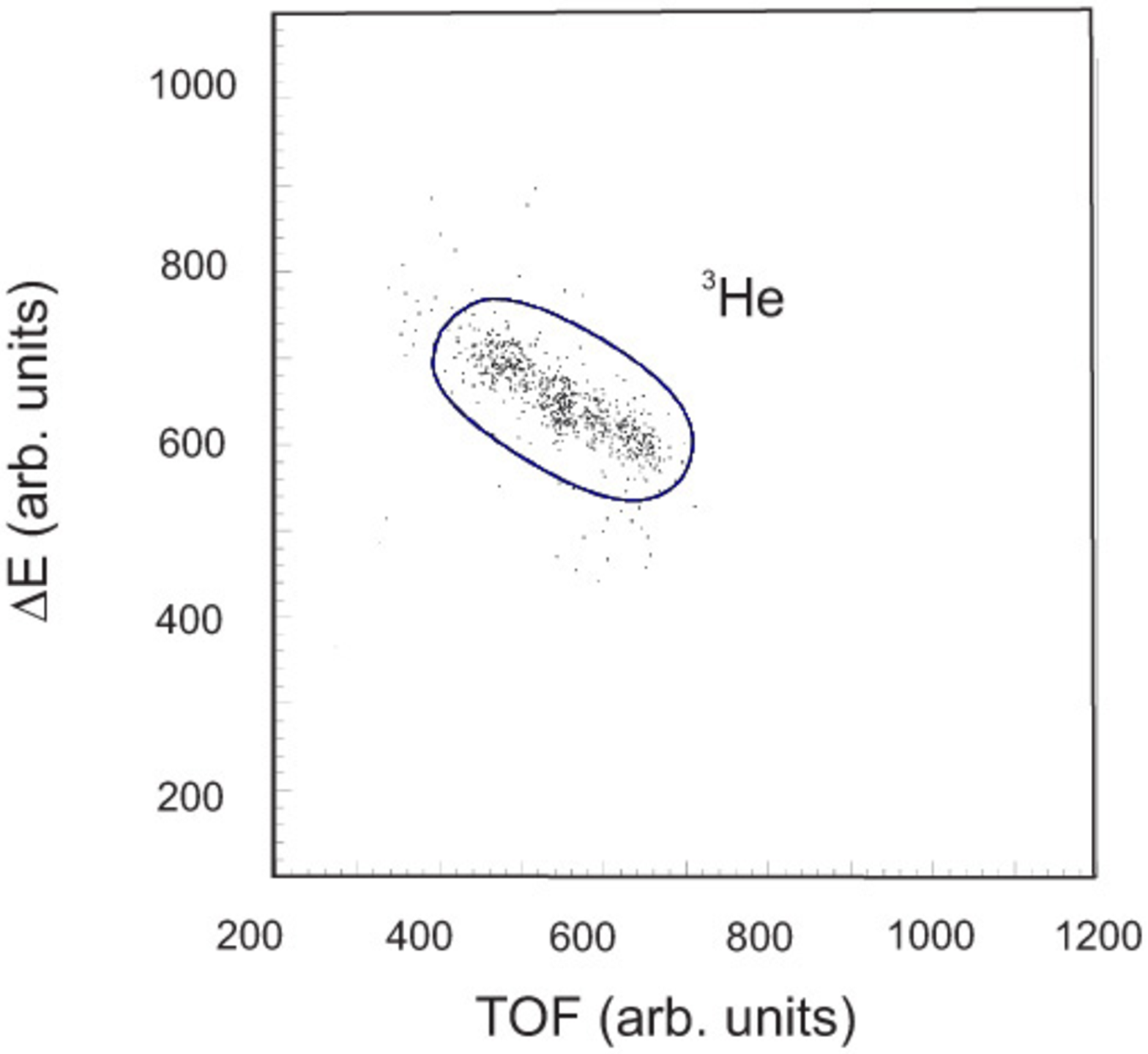}
\caption{Left panel: Particle identification in the focal plane of Big Karl
for proton-deuteron collisions at an excess energy of $Q=70$~MeV with respect
to the $^3$He$\,\pi^+\pi^-$ threshold. Events are plotted as function of the
energy loss in the first scintillator wall $\Delta E$ and the time of
flight (TOF) between the scintillator walls. The dominant proton events are
suppressed by imposing a threshold in the $\Delta E$ measurement. Right
panel: Same as Left but with the additional requirement of two hits in the
MOMO detector. This eliminates almost completely the triton and deuteron
events and confirms well the position and extent of the $^3$He band.}
\label{Fig:3He}
\end{center}
\end{figure}

The $^3$He ions produced close to threshold in the $dp(pd) \to{}
^3\textrm{He}\,\pi\pi$ reaction are confined to a small cone around the beam
direction and these were analyzed with the high resolution spectrograph Big
Karl~\cite{DRO1998}. Particle tracks were measured in the focal plane by two
planes of multi-wire drift chambers (MWDC), six chambers in each plane,
followed by two planes of scintillator walls. These walls allow particle
identification via $\Delta E-E$ as well as time-of-flight (TOF) measurements.
As is seen from Fig.~\ref{Fig:3He}, this combination led to the $^3$He being
well separated from tritons and deuterons even without requiring pion
detection in MOMO. The measurement of just the $^3$He yields the inclusive
cross section for $dp(pd) \to{} ^3\textrm{He}\,X$ reaction so that such data
would be comparable to those obtained in the initial ABC
experiments~\cite{ABA1963,BAN1973}. However, because the present experiments
were carried out at low excess energy, the unobserved state $X$ must
correspond to $\pi^+\pi^-$ or $\pi^0\pi^0$.

\begin{figure}[hbt]
\begin{center}
\includegraphics[width=0.7\columnwidth]{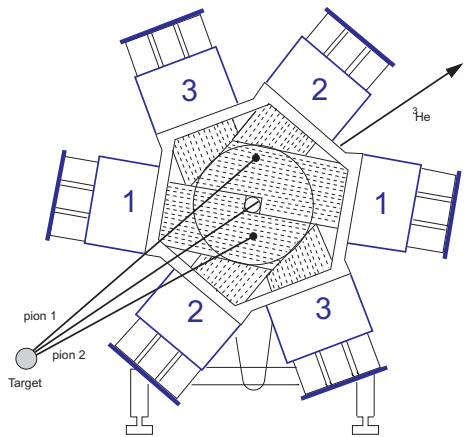}
\caption{(Color online) Front view of the MOMO vertex detector with the
indication of a typical event. Both the primary beam and the recoil $^3$He
detected in Big Karl pass through the central hole. The numbers denote the
different layers and the three boxes at the end of each read--out symbolize
the phototubes.} \label{Fig:vertexd}
\end{center}
\end{figure}

In order to reconstruct more completely the $pd(dp) \to{}
^3\textrm{He}\,\pi^+\pi^-$ events, the Big Karl spectrograph was supplemented
by the MOMO detector, which measured the two charged pions~\cite{BEL2007}.
MOMO consists of 672 scintillating fibers, arranged in three planes, denoted
by (1,2,3) in Fig.~\ref{Fig:vertexd}. The fibers are individually read out by
16-anode multichannel photomultipliers. The fibers in the three planes are
rotated by 60$^{\circ}$ with respect to each other and hits in three layers
are required in order to avoid combinatorial ambiguities. It is important to
note that the sign of the charge on each of the pions is not determined and
this automatically leads to the symmetrization of some of the differential
distributions.

\begin{figure*}[hbt]
\begin{center}
\includegraphics[width=0.7\textwidth]{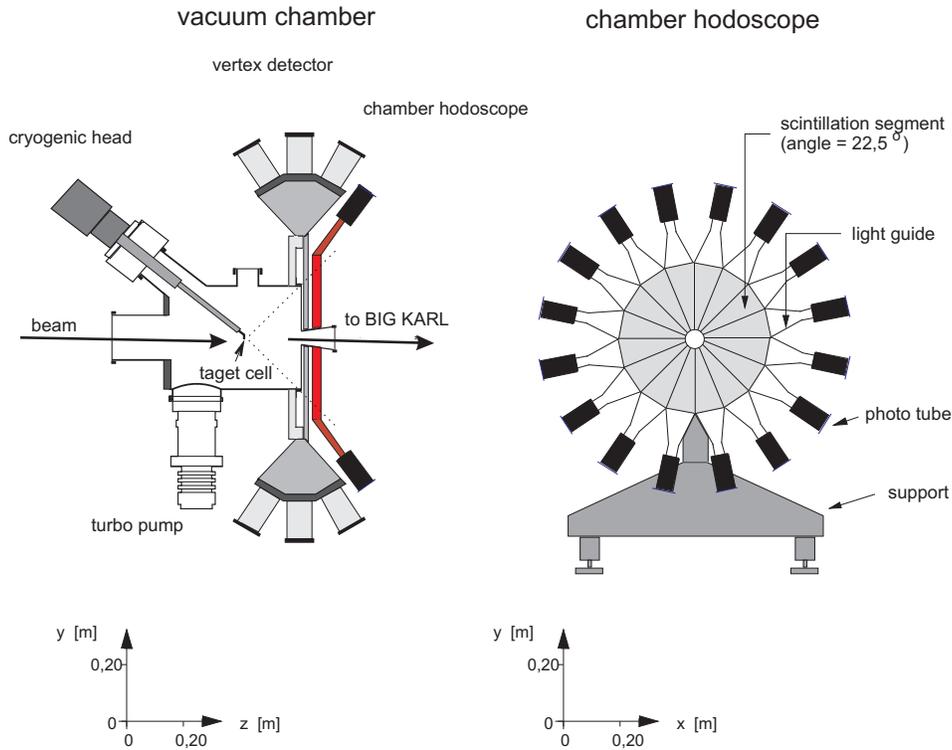}
\caption{(Color online) Left panel: Cross section through the target area
showing the location of the MOMO vertex hodoscope. The final wall in beam
direction is the segmented scintillator hodoscope (shown in red). Right
panel: View onto the segmented hodoscope placed after the MOMO detector.}
\label{Fig:momo_chamber}
\end{center}
\end{figure*}

The MOMO detector was placed perpendicular to the beam direction 20~cm
downstream of the target, outside a vacuum chamber, the end wall of which was
a 5~mm thick aluminum plate. The detector and its location are illustrated in
Fig.~\ref{Fig:momo_chamber}. The central hole, which subtended an angle of
6$^{\circ}$ at the target, allowed the passage of the primary beam and also
the $^3$He that were detected in Big Karl. The maximum angle of 45$^{\circ}$
was set by the physical dimensions of MOMO.

Each of the scintillating fibers is 2.5~mm thick but, when operating with a
deuteron beam, these were too thin to provide reliable energy information.
The MOMO wall was therefore complemented by a hodoscope consisting of 16
wedge-shaped 2~cm thick scintillators. This hodoscope, which is also shown in
Fig.~\ref{Fig:momo_chamber}, was already used in the study of the $pd\to
{^3\text{He}\,K^+K^-}$ reaction~\cite{BEL2007}.

The luminosity required to deduce absolute cross sections was measured in two
different ways. In the first method, applied in all runs, the luminosity was
measured with calibrated monitor counters placed in the forward hemisphere,
left and right of the target. During the calibration of the monitors, the
number of scattered particles was compared with the intensity of the direct
beam, as measured with scintillators in the beam exit of Big Karl. To avoid
dead-time effects in the hodoscope, the beam intensity was reduced by
de-bunching the beam between the ion source and the cyclotron injector. For
sufficiently small beam intensity the relation between monitors and hodoscope
is linear. In the actual production runs the counting rates in the monitors
was small enough to reduce the dead-time effects to a negligible level. The
systematic uncertainty in the beam intensity obtained using this procedure is
estimated to be 5\%. Combining this with a target thickness uncertainty, that
is also about 5\%, the total systematic uncertainty in the cross section
normalization is conservatively estimated to be below 10\%.

The results were controlled by a second method that is independent of the
target thickness. Elastic $pd$ or $dp$ scattering was studied with two
telescopes that measured protons and deuterons in coincidence. The
telescopes, each consisting of two silicon counters, were placed left and
right of the target at positions determined by elastic scattering kinematics.
The normalization was then deduced using the cross sections for elastic
proton-deuteron scattering taken from the compilation of Ref.~\cite{STE2002}.
The results of the two methods were consistent within error bars.

Although, unlike the CELSIUS experiments~\cite{BAS2006,AND2000}, there was no
$\pi^0$ detector, it was still possible to extract estimates for the $pd(dp)
\to{} ^3\textrm{He}\,\pi^0\pi^0$ cross section by comparing the inclusive
$pd(dp) \to{} ^3\textrm{He}\,X^0$ cross section deduced from the Big Karl
measurement with that for $pd(dp) \to{} ^3\textrm{He}\,\pi^+\pi^-$ obtained
from the combined Big Karl and MOMO data. However, such a subtraction does
depend on precise evaluations of the $\pi^+\pi^-$ acceptance in MOMO.

\begin{figure}[hbt]
\begin{center}
\includegraphics[width=0.9\columnwidth]{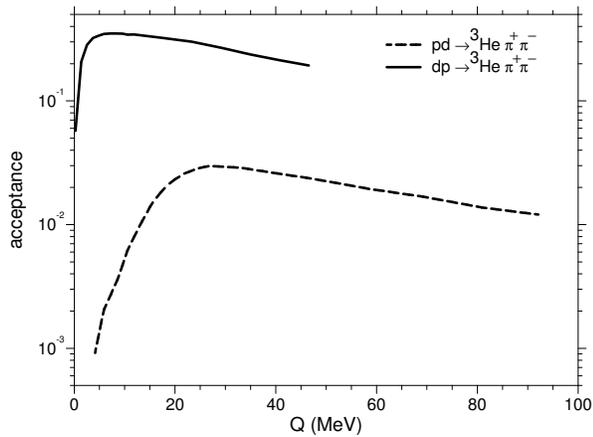}
\caption{Comparison of the acceptances of the full detection system for the
$pd \to{} ^3\textrm{He}\,\pi^+\pi^-$ and $dp \to{} ^3\textrm{He}\,\pi^+\pi^-$
reactions as functions of the excess energy $Q$.} \label{Fig:acceptance}
\end{center}
\end{figure}

The acceptance of the overall system for the measurement of the
$^3\textrm{He}\,\pi^+\pi^-$ final state is generally much higher for the
deuteron than the proton beam. Part of this is due to the tighter forward
cone of the $^3$He detected in Big Karl but there other important effects of
the forward momentum boost, in particular the higher probability that the
pions will emerge with angles below 45$^{\circ}$ and thus be detected in
MOMO. Decay losses are also less in inverse kinematics. The overall
acceptance estimates for the standard and inverse kinematics are presented in
Fig.~\ref{Fig:acceptance}  for the energy ranges relevant for the current
measurements.

The acceptance falls at very low $Q$ because of the beam-pipe hole shown in
Fig.~\ref{Fig:momo_chamber} but, away from this region, it decreases steadily
with increasing $Q$, though with the acceptance in inverse kinematics being
about an order of magnitude higher than with the proton beam. This factor is
not compensated by the differences in beam intensities, which were typically
$5\times10^8$ protons per spill of 4~s length and 11~s repetition rate and
$7\times10^9$ deuterons per spill of 30~s length. Measurements with the
proton beam are therefore severely limited for both low and high excess
energy.

%
%
\section{Experimental results}
\label{results}

\begin{figure}[h!]
\begin{center}
\includegraphics[width=1.0\columnwidth]{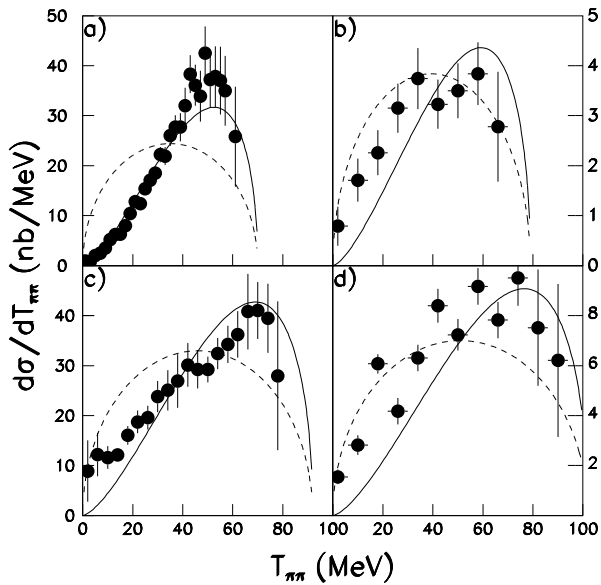}
\caption{MOMO measurements of the differential cross section for $pd \to{}
^3\textrm{He}\,\pi^+\pi^-$ at a) $Q=70$~MeV and c) $Q=92$~MeV and $pd \to{}
^3\textrm{He}\,\pi^0\pi^0$ at b) $Q=79$~MeV and d) $Q=101$~MeV as functions
of the excitation energy $T_{\pi\pi}$ in the $\pi\pi$ system. The dashed
curves are non-relativistic phase-space distributions normalized to the
integrated cross sections and the solid ones represent phase space multiplied
by a $T_{\pi\pi}$ factor and similarly normalized. } \label{Fig:7092}
\end{center}
\end{figure}

The previous MOMO measurement of $pd \to{} ^3\textrm{He}\,\pi^+\pi^-$ at
$Q=70$~MeV~\cite{BEL1999} is shown in Fig.~\ref{Fig:7092}(a) along with
analogous data obtained at $Q=92$~MeV in Fig.~\ref{Fig:7092}(b). The message
from the two data sets is similar; there is no sign of any ABC enhancement
and the shapes of the differential cross sections look much closer to phase
space weighted by the $\pi\pi$ excitation energy than pure phase space.

By comparing the inclusive data obtained just with the use of Big Karl with
those where there was also signals in the MOMO detector it was possible to
get the estimates of the $pd \to{} ^3\textrm{He}\,\pi^0\pi^0$ cross section
at $Q=79$~MeV and $Q=101$~MeV shown in Figs.~\ref{Fig:7092}(c) and (d),
respectively. The higher excess energies noted here are a consequence of the
pion mass differences. These data are typically an order of magnitude lower
than for charged pion production. This indicates that, although
$I_{\pi\pi}=0$ production is not negligible at these energies, the dominant
production must be in $I_{\pi\pi}=1$. The non-vanishing of the isovector
production was already evident in the direct measurements at CELSIUS at
$Q=28$~MeV~\cite{AND2000}.

Given that the $\pi^0\pi^0$ data were obtained by comparing two big numbers,
the associated error bars are much larger and it is less easy to make firm
conclusions regarding the shapes of the distributions. Nevertheless, there
does seem to be some tendency for the cross sections to be pushed to higher
$\pi\pi$ excitation energies than would be suggested by phase space.

\begin{figure}[hbt]
\begin{center}
\includegraphics[width=0.9\columnwidth]{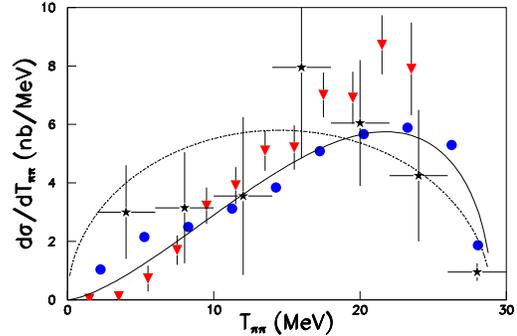}
\caption{(Color online) Cross section for the production of the
$^3\textrm{He}\,\pi^+\pi^-$ final state at an excess energy of $Q\approx
28$~MeV as a function of the excess energy $T_{\pi\pi}$ in the $\pi\pi$ rest
frame. The (blue) circles are MOMO data taken with a deuteron beam whereas
the (red) inverted triangles are the corresponding proton beam
data~\cite{BOH1998}. The CELSIUS data~\cite{AND2000} have been reduced by a
factor of 0.5 before being shown by the (black) stars. The chain curve is an
arbitrarily normalised phase space distribution and the solid curve is that
weighted with a $T_{\pi\pi}$ factor. } \label{Fig:28}
\end{center}
\end{figure}

Data on the $pd \to{} ^3\textrm{He}\,\pi^+\pi^-$ reaction had been obtained
at CELSIUS at $Q\approx 28$~MeV~\cite{AND2000}. In view of the limited
statistics in the CELSIUS experiment, we have used the MOMO detector to
explore this region with both proton and deuteron beams. All three data sets
are shown in Fig.~\ref{Fig:28} where, in order to compare the shapes of the
distributions, the CELSIUS results have been reduced by a factor of 0.5. This
factor is significant in comparison to the quoted 10\% statistical
uncertainty in the luminosity~\cite{AND2000}.

The shapes of the three data sets are broadly consistent. Any difference
between the MOMO $pd$ and $dp$ normalizations is not inconsistent with the
overall systematic uncertainties discussed earlier. However, it must be noted
that in $pd$ kinematics there is a loss of acceptance for very large
$\pi^+\pi^-$ excitation energies and no points are shown above about 24~MeV.

There is little sign of an ABC effect, \emph{i.e.}, any enhancement at low
$\pi\pi$ excitation energy $T_{\pi\pi}$, though the larger acceptance $dp$
data do show more strength in this region than the $pd$ results. Just as for
the original 70~MeV MOMO data, the results are better described by weighting
the phase space distribution by a $T_{\pi\pi}$ factor, as if the two pions
were emerging in a relative $p$ wave.

\begin{figure}[h!]
\begin{center}
\includegraphics[width=0.9\columnwidth]{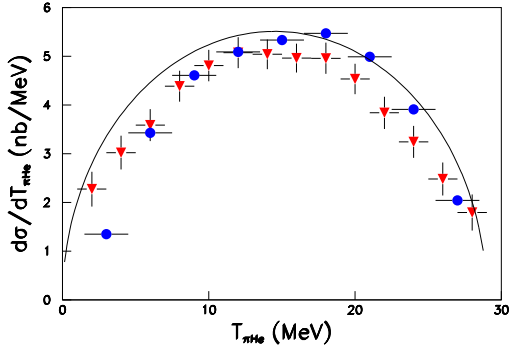}
\caption{(Color online) Cross section for the $dp \to{}
^3\textrm{He}\,\pi^+\pi^-$ (blue circles) and $pd \to{}
^3\textrm{He}\,\pi^+\pi^-$ (red inverted triangles) reactions at an excess
energy of $Q\approx 28$~MeV as a function of the excess energy
$T_{\pi{}^3\textrm{He}}$ in the $\pi^{3}$He rest frame. These MOMO data are
compared with an arbitrarily normalized non-relativistic phase space
distribution. } \label{Fig:28b}
\end{center}
\end{figure}

The distortion of phase space is far less evident in the distribution in the
$\pi{}^3$He energies shown in Fig.~\ref{Fig:28b}. Unlike the higher energy
data~\cite{BAS2006,MIE2016}, the lack of a magnetic field did not allow
separate plots to be made for $\pi^+$ and $\pi^-$.

There are no major discrepancies between the two MOMO data sets at
$Q=28$~MeV, which is some confirmation of the reliability of the MOMO
acceptance evaluations. Nevertheless, it must be assumed that the results
obtained with the deuteron beam are the more reliable because of the much
larger acceptance shown in Fig.~\ref{Fig:acceptance}.

\begin{figure}[h!]
\begin{center}
\includegraphics[width=1.1\columnwidth]{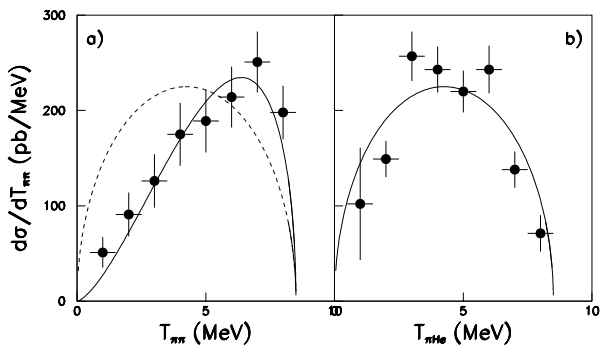}
\caption{MOMO measurements of the differential cross section for the $dp
\to{} ^3\textrm{He}\,\pi^+\pi^-$ reaction at $Q=8$~MeV in terms of a) the
excitation energy in the $\pi^+\pi^-$ system, and b) in the $\pi^3$He system.
The dashed curve in a) shows the shape of the phase-space distribution
whereas the solid one is phase space modified by a $T_{\pi\pi}$ factor. In
the $\pi^3$He system of b), only the phase-space shape is shown.}
\label{Fig:8}
\end{center}
\end{figure}

Exactly the same behavior is seen at $Q=8$~MeV as that commented upon at
28~MeV. Thus the $\pi^+\pi^-$ distribution shown in Fig.~\ref{Fig:8}a is well
described if the phase-space function is modified by a $T_{\pi\pi}$ factor.
On the other hand, the $\pi^3$He distribution of Fig.~\ref{Fig:8}b shows much
less deviation from phase space though this may, in part, be linked to this
being an average of the $\pi^+{}^3$He and $\pi^-{}^3$He spectra.

\begin{figure}[h!]
\begin{center}
\includegraphics[width=0.9\columnwidth]{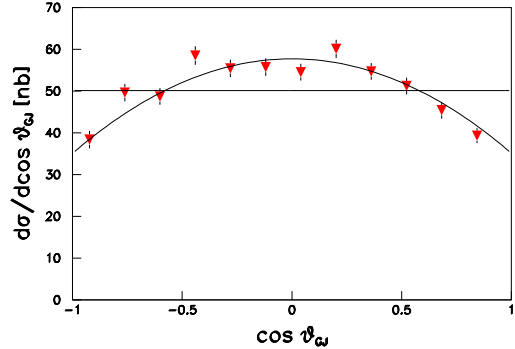}
\caption{(Color online) Distribution of the MOMO $dp \to{}
^3\textrm{He}\,\pi^+\pi^-$  data at $Q=28$~MeV in the Gottfried-Jackson
angle. The data are symmetric about 90$^{\circ}$ because the sign of the
charges on the pions was not measured. The curve shown,
$\dd\sigma/\dd\!\cos\theta_{GJ} =57.7 -22.6\cos^2\theta_{GJ}$, is a best fit
to the data assuming a linear dependence in $\cos^2\theta_{GJ}$. }
\label{Fig:28c}
\end{center}
\end{figure}

\begin{figure}[h!]
\begin{center}
\includegraphics[width=1.0\columnwidth]{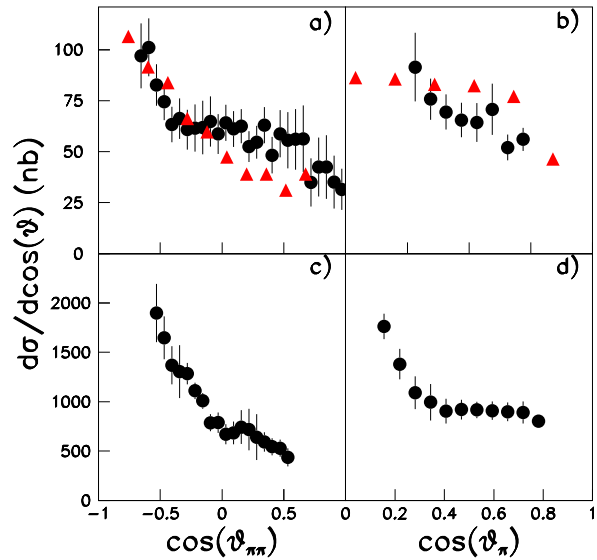}
\caption{(Color online) Differential cross section for the $dp \to{}
^3\textrm{He}\,\pi^+\pi^-$ at excess energies of $Q = 28$~MeV (a+b) and
$Q=92$~MeV (c+d) in terms of the pion opening angle $\theta_{\pi\pi}$ and the
angle $\theta_{\pi}$ between the outgoing pion and the incoming beam
direction, both angles being evaluated in the overall c.m.\ frame. The
(black) circles were taken in $pd$ kinematics but at $Q=28$~MeV data (blue
triangles) were also obtained in $dp$ kinematics with a much enhanced
acceptance.} \label{Fig:287092}
\end{center}
\end{figure}

The clearest proof for the importance of higher partial waves in the $dp
\to{} ^3\textrm{He}\,\pi^+\pi^-$ reaction even as close to threshold as
$Q=28$~MeV is provided by the distribution in the Gottfried-Jackson angle
$\theta_{GJ}$~\cite{GOT1964}. This is the angle between the relative momentum
between the two pions and the direction of the deuteron beam, evaluated in
the dipion rest frame. Any anisotropy here is a signal for higher partial
waves in the $\pi^+\pi^-$ system. The MOMO data shown in Fig.~\ref{Fig:28c}
are symmetric about 90$^{\circ}$ because the $\pi^+$ and $\pi^-$ are not
distinguished in this detector. The clear deviation from isotropy proves that
the dipion cannot be in a pure $s$ wave. Such a behavior could be a signal
for a superposition of $s$- and $p$-wave pion pairs but higher partial waves
are not definitively excluded. The sign of the $\cos^2\theta_{GJ}$ term is
opposite to that we found for $K^+K^-$ production~\cite{BEL2007}, though this
could be influenced by $\phi$ production.

Other angular distributions can be derived from the MOMO data and we show in
Fig.~\ref{Fig:287092} those with respect to the $\pi^+\pi^-$ opening angle,
$\theta_{\pi\pi}$, and one pion with respect to the beam direction,
$\theta_{\pi}$, both in the overall CM frame. At $Q=28$~MeV data were
obtained in both the original $pd$ kinematics and also with the much
increased acceptance offered by $dp$ kinematics. The biggest disagreement
between the 28~MeV results obtained with the two kinematics is at large
$\cos\theta_{\pi\pi}$ in Fig.~\ref{Fig:287092}a. This is the region
preferentially associated with small $T_{\pi\pi}$ and we already saw a
similar discrepancy in Fig.~\ref{Fig:28}.

\begin{figure}[h!]
\begin{center}
\includegraphics[width=1.0\columnwidth]{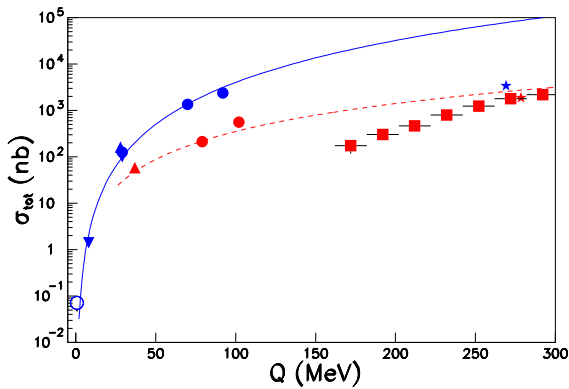}
\caption{(Color online) Dependence of the total cross sections for $pd(dp)
\to{} ^3\textrm{He}\,\pi^+\pi^-$ (blue) and $pd(dp) \to{}
^3\textrm{He}\,\pi^0\pi^0$ (red) on the excess energy $Q$. The curves are
arbitrarily normalized $Q^3$ and $Q^2$ shapes for $\pi^+\pi^-$ and
$\pi^0\pi^0$ production, respectively. The closed circles represent MOMO $pd$
data whereas those taken in $dp$ are shown as inverted triangles. The
triangles are low energy CELSIUS points~\cite{AND2000} and the stars are high
energy CELSIUS-WASA points obtained in $pd$ kinematics~\cite{BAS2006},
renormalized by a factor of 1.5~\cite{RIO2014}. The squares represent $pd
\to{} ^3\textrm{He}\,\pi^0\pi^0$ data obtained in $dd$ collisions within a
spectator model~\cite{ADL2015,RIO2014}. The near-threshold IUCF
measurement~\cite{BET1996} is indicated by an open circle. It should be noted
that the data points cannot be distinguished for $Q\approx 28$~MeV.}
\label{Fig:sigtot}
\end{center}
\end{figure}

Further evidence for the anomalous behavior of the $pd(dp) \to{}
^3\textrm{He}\,\pi\pi$ reaction at low energies is to be found in the
variation of the total cross section with $Q$ that is shown in
Fig.~\ref{Fig:sigtot}. A simple $Q^2$ phase-space dependence describes well
the $pd(dp) \to{} ^3\textrm{He}\,\pi^0\pi^0$ data but near threshold the
$Q^3$ dependence seen for $pd(dp) \to{} ^3\textrm{He}\,\pi^+\pi^-$ must
reflect the presence of higher partial waves. However, at $Q\approx 270$~MeV,
where the ABC enhancement is obvious~\cite{BAS2006}, the $Q^3$ dependence
must have moderated considerably. This suggests that there might be some
$I_{\pi\pi}=1$ contribution that is important at low $Q$ that becomes less
significant at high $Q$. This conclusion is consistent with the CELSIUS
isospin decomposition at low energy~\cite{AND2000}.

Values of the $pd \to{} ^3\textrm{He}\,\pi^0\pi^0$ total cross section were
also obtained from measurements of the $dd\to{} n_{\rm sp}{}
^3\textrm{He}\,\pi^0\pi^0$ reaction, assuming that the unobserved neutron to
be a true \emph{spectator}. By measuring the $^3$He and the two $\pi^0$, the
reaction could be studied over a wide $Q$ range while using a fixed deuteron
beam energy of 1.7~GeV~\cite{ADL2015,RIO2014}. The energy dependence
indicated by these points shown in Fig.~\ref{Fig:sigtot} seems to be at odds
with the data at lower $Q$ but it must be stressed that this conclusion does
depend on the use of the spectator model for large Fermi momenta.

%
%
\section{Interpretation}
\label{interpretation}

There is no universally accepted model for the ABC effect in the $pd
\to{}^3\textrm{He}\,X^0$ reaction but it is clear from all the data shown in
Sec.~\ref{results} that the possible presence of an ABC effect depends
strongly upon the excess energy. Below $Q\approx 100$~MeV there is no sign of
any ABC enhancement.

It has been argued that the ABC effect is closely associated with the decay
of the $d^{\star}(2380)$ dibaryon resonance in $np\to
d\pi^0\pi^0$~\cite{BAS2009} and that this resonance might also play an
important role in more complicated reactions, such as $pd
\to{}^3\textrm{He}\,X^0$~\cite{ADL2012}. Although this does offer a natural
explanation for the strong energy dependence of the ABC production, the
momentum transfers seem to be very large for a model involving a
$d^{\star}(2380)$ and a spectator nucleon.

There is good evidence that at high $Q$ the ABC effect is dominantly
isoscalar in character. On the other hand, at $Q=28$~MeV the production of
isovector pion pairs is the larger and, at our two highest energies,
isoscalar production, though small, is certainly non-zero. As a consequence
the $\pi^+\pi^-$ pair cannot be purely in a relative $p$-wave with
$I_{\pi\pi}=1$, as we assumed earlier when describing our $Q=70$~MeV
data~\cite{BEL1999}.

Nevertheless, the $\pi^+\pi^-$ data for $Q<100$~MeV could still be described
in terms of a dominant $p$-wave plus a small amount of $s$-wave that is
required by the $\pi^0\pi^0$ data of Fig.~\ref{Fig:7092}. This would still
yield an energy dependence of the total cross section that is close to the
$Q^3$ fit shown in Fig.~\ref{Fig:sigtot}. One difficulty with this assumption
is to be found in the shapes of the $\pi^0\pi^0$ spectra shown in
Figs.~\ref{Fig:7092}b and \ref{Fig:7092}d. Though the uncertainties here are
large, due to the subtraction of the $\pi^+\pi^-$ data from the inclusive
spectra, they seem to show features that are similar to the $\pi^+\pi^-$
distributions, with a preference to higher $T_{\pi\pi}$ values than those
suggested by phase space. This is what might be expected in a two-step
model~\cite{FAL2000}.

\begin{figure}[h!]
\begin{center}
\includegraphics[width=1.0\columnwidth]{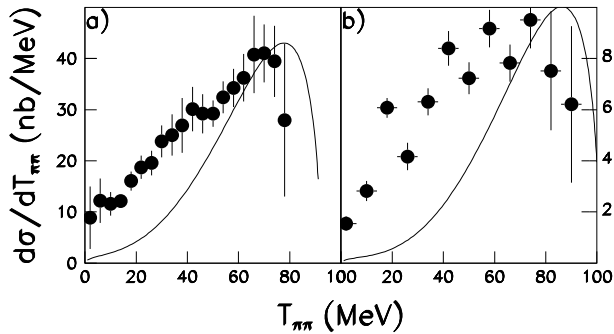}
\caption{a) Differential cross section for $pd \to{}
^3\textrm{He}\,\pi^+\pi^-$ at $Q=92$~MeV compared to the predictions of the
two-step model~\cite{FAL2000}. b) Differential cross section for $pd \to{}
^3\textrm{He}\,\pi^0\pi^0$ at $Q=101$~MeV obtained by comparing data with and
without signals in the MOMO detector. The theoretical predictions for
$\pi^0\pi^0$ production are reduced by a factor of $0.3$.} \label{Fig:92}
\end{center}
\end{figure}

A classical two-step model was first proposed for $\eta$ production in the
$pd \to{}^3\textrm{He}\,\eta$ reaction~\cite{KIL1991} and this was later put
on a quantum mechanical basis~\cite{FAL1995}. When applied to two-pion
production, it is assumed that the reaction consists of pion production
through $pp\to d\pi^+$ followed by $\pi^+n\to \pi^+\pi^-(\pi^0\pi^0)p$, with
the final proton and deuteron fusing to form the observed
$^3$He~\cite{FAL2000}. As currently implemented, only the contribution from
isoscalar pion pairs has been estimated as a function of the excitation
energy in the $\pi\pi$ system. The predictions of the model for the
differential distributions at the highest MOMO energy are compared with the
experimental data in Fig.~\ref{Fig:92}, where the normalization of the form factors
is determined from the threshold rate of the $pd \to{}^3\textrm{He}\,\eta$
reaction~\cite{FAL1995}.

The curves are both pushed towards the maximum $T_{\pi\pi}$ but, since this
corresponds to isoscalar pion pairs, it is not due to pion $p$ waves but it
is rather a feature of the $\pi^+n\to \pi^+\pi^-(\pi^0\pi^0)p$ amplitude,
which was taken from the Valencia model~\cite{VAC1999}. This striking
behavior is due to a cancelation at low $\pi\pi$ excitation energies between
a contact term and the contribution from the Roper resonance. The model was
tuned to fit the $\pi^-p\to \pi^0\pi^0n$ experimental data in the low $Q$
region and it is not valid to continue it to higher energies to investigate
the ABC phenomenon. Despite its failings at low $T_{\pi\pi}$, the model
predicts the right order of magnitude for $\pi^+\pi^-$ production, though the
predictions have to be reduced by a factor of $0.3$ in order to describe the
$\pi^0\pi^0$ data.

\begin{figure}[h!]
\begin{center}
\includegraphics[width=1.0\columnwidth]{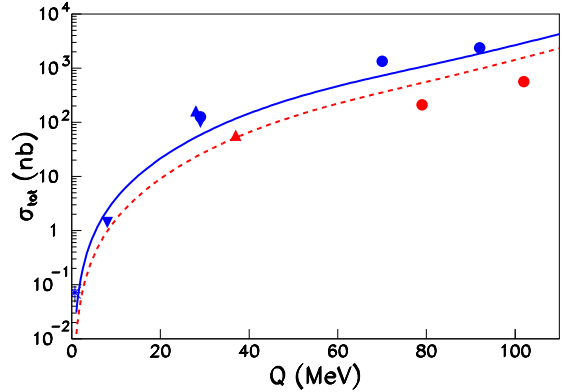}
\caption{(Color online) The low energy data of Fig.~\ref{Fig:sigtot}, showing
total cross sections for the $pd \to{} ^3\textrm{He}\,\pi^+\pi^-$ and $pd
\to{} ^3\textrm{He}\,\pi^0\pi^0$ reactions. These are compared with the
predictions of the two-step model of Ref.~\cite{FAL2000} for isoscalar
$\pi\pi$ pairs. The solid (blue) curve is for $\pi^+\pi^-$ production and the
dashed (red) curve for $\pi^0\pi^0$. } \label{Fig:sigtot3}
\end{center}
\end{figure}

The predictions of the energy dependence of the total cross sections for
isoscalar pion pair production in the $pd \to{} ^3\textrm{He}\,\pi^+\pi^-$
and $pd \to{} ^3\textrm{He}\,\pi^0\pi^0$ reactions are shown in
Fig.~\ref{Fig:sigtot3}. Given the uncertainty in the model and the fact that
only the $I_{\pi\pi}=0$ contribution is predicted, the estimate of the $pd
\to{} ^3\textrm{He}\,\pi^+\pi^-$ total cross section is reasonable. The same
cannot be said for the $pd \to{} ^3\textrm{He}\,\pi^0\pi^0$ prediction.
Though it is close to the value obtained by the CELSIUS group at
37~MeV~\cite{AND2000}, the curve is over three times too high compared to the
MOMO data at 79 and 101~MeV. The MOMO values, of course, result from indirect
measurements, so that systematic uncertainties may be large.

The only way that a factor of ten between the $\pi^+\pi^-$ and $\pi^0\pi^0$
production cross sections could arise is if the $I_{\pi\pi}=1$ production
were very much stronger than $I_{\pi\pi}=0$. If this proves to be the case,
the two-step model must have given a gross overestimate of the $I_{\pi\pi}=0$
contribution to $\pi^+\pi^-$ production.

%
%
\section{Conclusions}
\label{conclusions}

New data have been presented on both the $pd\to{} ^3\textrm{He}\,\pi^+\pi^-$
and $dp\to{} ^3\textrm{He}\,\pi^+\pi^-$ reactions at excess energies $Q <
100$~MeV, where the $^3$He was measured in a high resolution spectrograph and
the charged pions in the MOMO vertex detector. Though the results obtained
are generally consistent, the acceptance of the whole system is much higher
with the deuteron beam and these results are much to be preferred. In all
cases the differential cross sections seemed suppressed at low $M_{\pi\pi}$
invariant masses compared to phase space and there was certainly no sign in
the $\pi^+\pi^-$ spectrum of the ABC enhancement that is so prevalent in
higher energy data.

Though, as we previously reported~\cite{BEL1999}, the data could be an
indication of isovector $\pi^+\pi^-$ $p$-waves, there are other possible
explanations and the behavior could be governed by that present in the
$\pi^-p\to \pi^0\pi^0 n$ amplitudes, where $p$-waves are forbidden. Such a
model does reproduce features of the observed mass distributions but it would
have to be extended to include both $I_{\pi\pi}=1$ contributions and angular
distributions before it could be considered a satisfactory theory. Of
particular importance in this regard is the distribution in the
Gottfried-Jackson angle, where our data clearly prove that there must be
contributions from higher partial waves in the $\pi\pi$ system at energies
even as low as $Q=28$~MeV.

The comparison of data taken with and without a charged pion signal in MOMO
allowed estimates to be made for the $pd\to{} ^3\textrm{He}\,\pi^0\pi^0$
production rates. The systematic uncertainties are, of course, larger and
direct measurements, such as those achieved with WASA~\cite{BAS2006}, should
also be attempted. The comparison of the current MOMO $\pi^+\pi^-$ and
$\pi^0\pi^0$ data for $Q>70$~MeV can only be understood if the pion pairs are
overwhelmingly produced with $I_{\pi\pi}=1$.

For $Q\gtrsim 180$~MeV there is a strong ABC effect whereas for $Q\lesssim
100$~MeV the ABC is completely absent. Data are sadly lacking in the
intermediate energy interval to show how the ABC develops between 100 and
180~MeV. The only quality data that exist in this region were taken in
deuteron-deuteron collisions~\cite{ADL2015,RIO2014} and they rely on the
spectator model being valid at large Fermi momenta. The situation can only be
clarified by measurements of free $dp \to{} ^3\textrm{He}\,\pi^+\pi^-$ and
$dp \to{} ^3\textrm{He}\,\pi^0\pi^0$ reactions in this energy range.

%
%
\section*{ACKNOWLEDGEMENTS}
We wish to thank the COSY machine crew for providing the high quality proton
and deuteron beams necessary for these experiments. We are also indebted to
the Big Karl technical staff for their tireless efforts. Correspondence with
Dr Perez del Rio, who provided the numerical values of the quasi-free data
used in Fig.~\ref{Fig:sigtot}, has been most helpful. Support by
Forschungszentrum J\"{u}lich (FFE) and Bundesministerium f\"{u}r Bildung und
Forschung (BMBF) is gratefully acknowledged.
%
%

\end{document}